\pdfoutput=1 
\documentclass{JINST}

\usepackage[squaren,textstyle]{SIunits}
\usepackage{subfig}
\newcommand{\micron}{\ensuremath{\micro\rm{m}}}

\title{\boldmath Results of the 2015 testbeam of a 180 nm AMS High-Voltage CMOS sensor prototype}

\author{
	\mbox{M. Benoit$^a$}, 
	\mbox{J. Bilbao de Mendizabal$^a$},
	 \mbox{G. Casse$^b$}, 
	 \mbox{H. Chen$^c$}, 
	 \mbox{K. Chen$^c$}, 
	 \mbox{F.A. Di Bello$^{a,1}$}, 
	 \mbox{D. Ferrere$^a$}, 
	 \mbox{T. Golling$^a$}, 
	 \mbox{S. Gonzalez-Sevilla$^{a,1}$}, 
	 \mbox{G. Iacobucci$^a$}, 
	 \mbox{F. Lanni$^c$}, 
	 \mbox{H. Liu$^{c,d}$}, 
 	 \mbox{F. Meloni$^e$}, 
	 \mbox{L. Meng$^{a,b}$}, 
	 \mbox{A. Miucci$^{a,2}$}, 
	 \mbox{D. Muenstermann$^{a,3}$}, 
	 \mbox{M. Nessi$^{a,f}$}, 
	 \mbox{I. Peri\'c$^g$}, 
	 \mbox{M. Rimoldi$^e$}, 
	 \mbox{B. Ristic$^{a,f}$}, 
	 \mbox{M. Vicente Barrero Pinto$^a$}, 
	 \mbox{J. Vossebeld$^b$}, 
	 \mbox{M. Weber$^e$}, 
	 \mbox{W. Wu$^c$},
	  \mbox{and L. Xu$^c$}\\
\llap{$^a$}D\'epartement de Physique Nucl\'eaire et Corpusculaire (DPNC), University of Geneva, 24 quai Ernest Ansermet 1211 Gen\`eve 4, Switzerland\\
\llap{$^b$}Department of Physics, University of Liverpool, The Oliver Lodge Laboratory, Liverpool L69 7ZE, UK\\
\llap{$^c$}Brookhaven National Laboratory (BNL), P.O. Box 5000, Upton, NY 11973-5000, USA\\
\llap{$^d$}Department of Modern Physics, University of Science and Technology of China, Hefei, Anhui 230026, China\\
\llap{$^e$}Albert Einstein Center for Fundamental Physics and Laboratory for High Energy Physics, University of Bern, Sidlerstrasse 5, CH-3012 Bern, Switzerland\\
\llap{$^f$}European Organization for Nuclear Research (CERN), 385 route de Meyrin, 1217 Meyrin, Switzerland\\
\llap{$^g$}Karlsruhe Institute of Technology (KIT), Kaiserstra{\ss}e 12, 76131 Karlsruhe, Germany\\ \\
E-mail: \email{Sergio.Gonzalez@unige.ch}
}

\footnotetext[1]{Corresponding author}
\footnotetext[2]{Now at Albert Einstein Center for Fundamental Physics and Laboratory for High Energy Physics, University of Bern, Sidlerstrasse 5, CH-3012 Bern, Switzerland}
\footnotetext[3]{Now at Physics Department, Lancaster University, Lancaster, UK, LA1 4YB.}

\abstract{Active pixel sensors based on the High-Voltage CMOS technology are being investigated as a viable option for the future pixel tracker of the ATLAS experiment at the High-Luminosity LHC. This paper reports on the testbeam measurements performed at the H8 beamline of the CERN Super Proton Synchrotron on a High-Voltage CMOS sensor prototype produced in 180 nm AMS technology. 
Results in terms of tracking efficiency and timing performance, for different threshold and bias conditions, are shown.
}


\begin{document}
\maketitle
\flushbottom

\section{Introduction}
\label{sec:intro}

Active pixel sensors based on the High-Voltage CMOS (HV-CMOS) technology~\cite{ivan} are promising candidates for the next generation of pixel tracking detectors in High Energy Physics experiments. The operating principle is based on a thin {\em p}-type silicon bulk containing a deep {\em n}-well that implements CMOS pixel electronics ({\em e.g.} preamplifier, discriminator). A reverse bias voltage is applied to the diode formed by {\em n}-well with respect to the substrate to create a depletion region. The charge carriers, created by ionization, drift towards the {\em n}-well that acts as a collecting electrode. As the transistors are physically isolated from the substrate, a relatively high bias voltage ($\sim$100 V) can be applied to enlarge the depletion zone. A readout ASIC coupled to the HV-CMOS sensor is used to collect the signal and perform further digital processing.

This paper reports on the testbeam results obtained for a HV-CMOS sensor prototype produced in 180 nm technology by ams AG~\cite{ams} (AMS). This manuscript is organized as follows. Section~\ref{sec:setup} describes the experimental setup and the telescope used for track reconstruction. Section~\ref{sec:dut} describes the Device Under Test (DUT). The offline analysis framework is briefly explained in Sec.~\ref{sec:offline}. Finally, Sec.~\ref{sec:results} presents the main results.

\section{Experimental setup}
\label{sec:setup}

The AMS HV-CMOS detector prototype was tested in November 2015 at the H8 beamline of the CERN Super Proton Synchrotron (SPS), that provided a 180 GeV $\pi^+$ beam. A dedicated beam telescope (hereafter called FE-I4 telescope) was used to provide a standalone measurement of the trajectories of the incoming particles independently of the device under test.
	
\begin{figure}[tbp]
\centering
\includegraphics[width=0.95\textwidth]{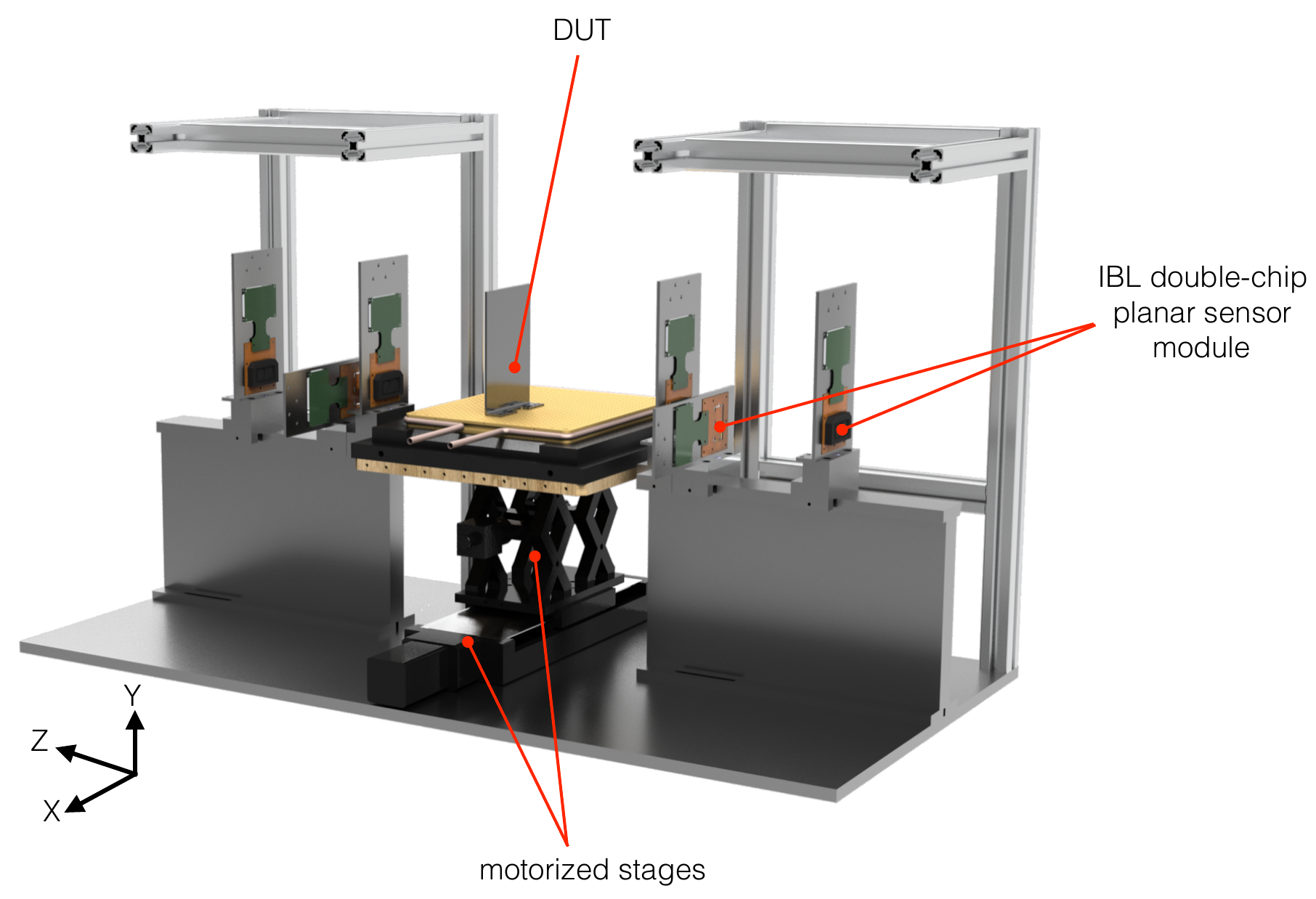}
\caption{A rendered view of the FE-I4 beam telescope. The coordinate system is defined with the $Z$-axis along the beam-axis, the $Y$-axis pointing vertically upwards and the $X$-axis pointing horizontally.}
\label{fig:telescope}
\end{figure}

\subsection{The FE-I4 beam telescope}
\label{sec:telescope}

The FE-I4 telescope~\cite{telescopePaper} consists of six ATLAS Insertable B-Layer (IBL)~\cite{ibl} double-chip (DC) pixel modules~\cite{ibl-modules} arranged in two arms of three modules each (see Fig.~\ref{fig:telescope}). Within a given arm, the first and last planes are installed vertically, with the most precise pixel coordinate (50~\micron-pitch, see below) along the $Y$-axis. The module in the middle is rotated by 90\degree\ to improve the spatial resolution along the $X$-direction. 

Each DC module uses a 200~\micro m-thick n$^+$-in-n planar silicon sensor produced by CiS~\cite{cis}. The detector is segmented in a matrix of \mbox{336 rows $\times$ 160 columns} with a pixel size of 50 $\times$ 250~\micron\squared. The p$^+$ back-side implant used for the high-voltage bias is surrounded by 13 guard rings shifted under the edge pixels, with a resulting total inactive area of only 200~\micron\ from the sensor edge. Two FE-I4B~\cite{fei4} ASICs produced in IBM 130 nm CMOS technology are bump-bonded to the detector for the signal readout. Although the sensor can be readout by two ASICs, only one of them was active during data-taking. The chip active area of \mbox{20.2 $\times$ 16.8~mm\squared} holds 26'880 identical \mbox{50 $\times$ 250~\micron\squared} pixels organized in a matrix of \mbox{336 rows $\times$ 80 columns}. Each pixel cell contains a two-stage amplifier followed by a discriminator with adjustable threshold, so that collected charge is translated into a 4-bit Time-over-Threshold (ToT) value measured in units of 25 ns (Bunch Crossing units or BC, corresponding to the LHC bunch spacing). The digital architecture is implemented such that four analog pixels share one common  logic cell. The hit information is stored locally at the pixel level until reception of a trigger signal. For test and calibration purposes, a test charge with selectable amplitude can be directly injected into each pixel pre-amplifier input. A threshold tuning Digital-to-Analog Converter (DAC) is used to apply a per-pixel threshold correction. At this testbeam, the discriminator thresholds of the telescope modules FEs were tuned to a charge of 2~$\rm{ke^-}$, with a final threshold dispersion of 70 $\rm{e^-}$. The ToT to charge conversion was tuned for each pixel to 9 ToT units for a deposited charge of 16~$\rm{ke^-}$.

\subsection{Data Acquisition System}

The Data Acqusition System (DAQ) of the FE-I4 telescope is based on the Reconfigurable Cluster Element (RCE), a generic computational element for DAQ systems developed at SLAC and based on System-on-Chip (SoC) technology. An Advanced Telecommunication Computing Architecture (ATCA) crate integrates several RCEs that are connected via optical links to a High Speed Input/Output (HSIO) development platform. The HSIO is a custom-made board  based on the Xilinx Virtex-4~\cite{xilinx} FPGA developed by SLAC as a generic DAQ interface for both ATLAS Pixel and Strip detectors. 

\subsection{Trigger}

The FE-I4B ASIC can provide a {\em HitBus} signal corresponding to the logical OR of selected pixel discriminator outputs. In the testbeam, the trigger signal for event acquisition is given by the coincidence (within one BC) of the {\em HitBus} signals of the first and last telescope modules. A pixel mask applied to the FE-I4B allows a precise definition of a Region-Of-Interest (ROI) on the triggering planes. By setting the ROI size to slightly exceed that of the device under test the data acquisition rate is optimized. A trigger rate of $\sim$18 kHz has been measured with the telescope in standalone mode (without DUT), and of  $\sim$6 kHz with up to two FE-I4-based DUTs~\cite{telescopePaper}.

\section{Device under test}
\label{sec:dut}

\begin{figure}[!htbp]
\centering
\subfloat[]{\label{fig:ccpd_v4}
\includegraphics[width=0.45\textwidth]{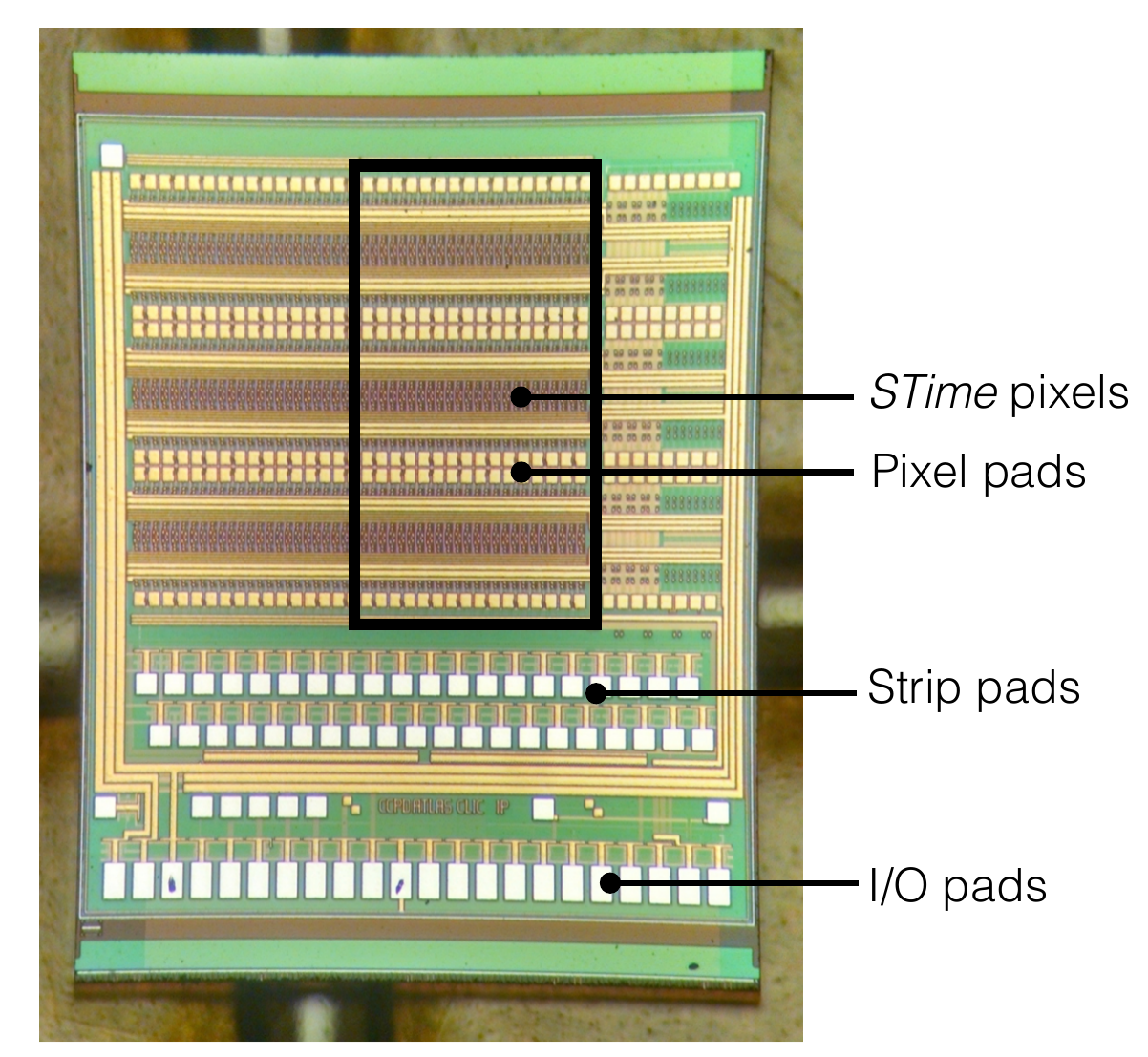}}
\hspace{0.5cm}
\subfloat[]{\label{fig:ccpd_fei4}
\includegraphics[width=0.45\textwidth]{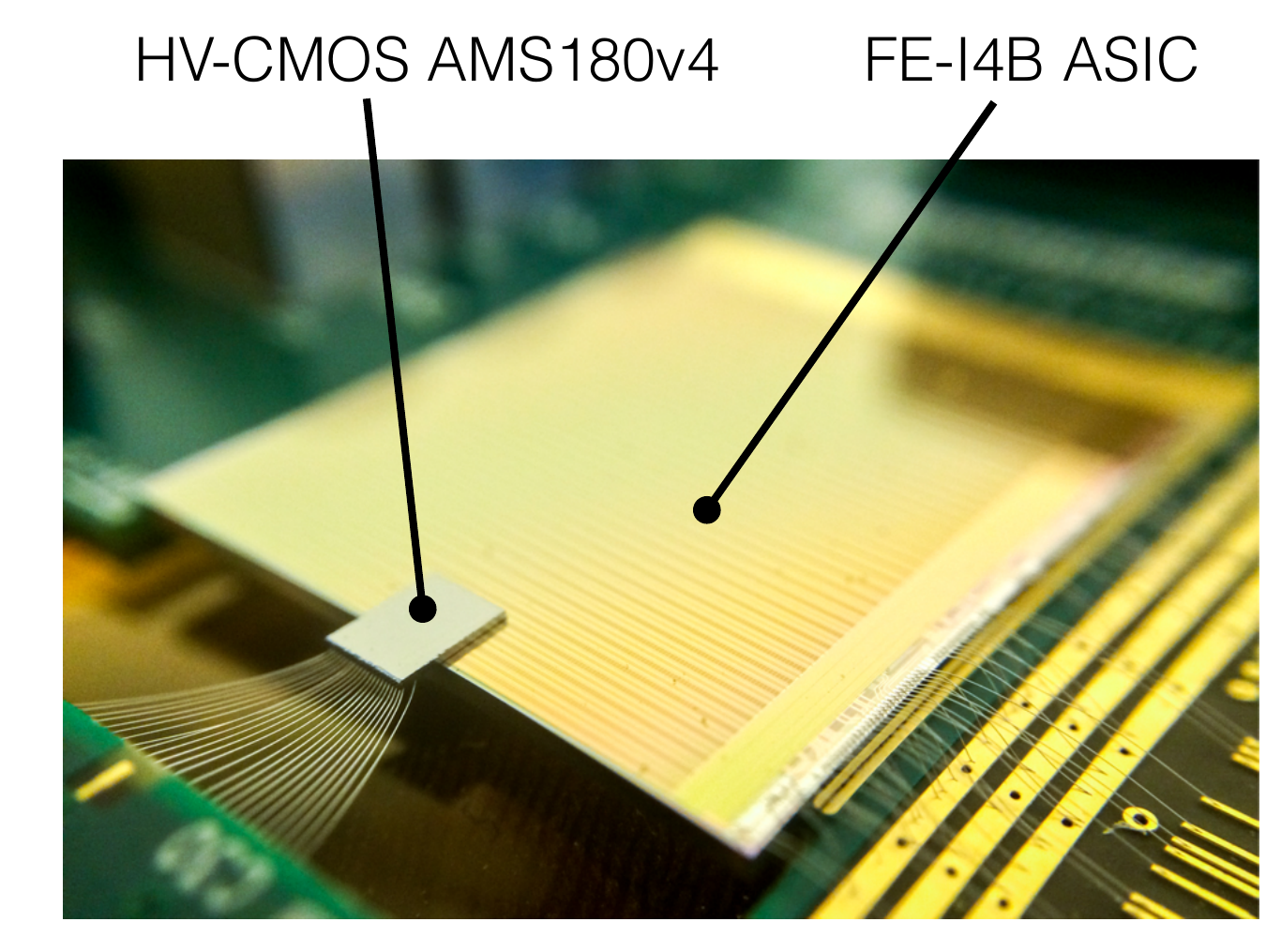}}
\caption{(a) AMS HV-CMOS 180 nm (version 4) bare sensor. The {\em I/O pads} are used to supply different analog voltages, the high-voltage bias and the digital inputs for the serial register configuration. The {\em Strip pads} were unused (these are needed only when operating the sensor with a strip readout chip). The {\em Pixel pads} have the same pitch as the FE-I4B pixels. (b) Final assembly of a FE-I4B pixel readout chip to a HV-CMOS v4 sensor via capacitive coupling.}
\label{fig:dut}   
\end{figure}

The DUT, referred as AMS180v4, is the fourth version of a series of prototypes produced in the HV-CMOS technology by the AMS foundry. The AMS180v4 has been produced in 180 nm with the AMS H18 process. The unit cell structure contains six pixels arranged in two columns and three rows, each pixel having an area of \mbox{$125 \times 33$~\micron\squared}. The substrate resistivity is $20~\Omega\cdot\rm{cm}$. The sensor (see Fig.~\ref{fig:ccpd_v4}) has a total size of \mbox{$\sim 2.4 \times 2.9$~mm\squared}, a thickness of 250~\micron\ and  it implements various pixel types. The results presented in this paper correspond to the baseline flavour, so-called {\em STime}, where each pixel implements a two-stage amplifier followed by a discriminator with tuneable output amplitude. Though the discriminator threshold is common to all pixels in the sensor, a 4-bit DAC (TuneDAC or TDAC) is used to provide a per-pixel threshold dispertion correction.

The sensor has been designed to be compatible with either a pixel or a strip readout chip. For this measurement, a FE-I4B pixel readout ASIC was capacitively coupled to the HV-CMOS sensor by means of a $< 1$~\micron-thick epoxy glue layer (see Fig.~\ref{fig:ccpd_fei4}). The SET ACC\micro RA 100 flip-chip bonder~\cite{set} was used to bond the assembly with a post-bonding accuracy of 1~\micron\ and good coplanarity (maximum deviation of 100 nm over a total length of 2.9 mm). After assembly, both the HV-CMOS sensor and the FE-I4B ASICs are wire-bonded to auxiliary boards (see section~\ref{sec:dut_threshold}), providing all input voltages and control lines for both ICs. 

\begin{figure}[tbp]
\centering
\includegraphics[width=\textwidth]{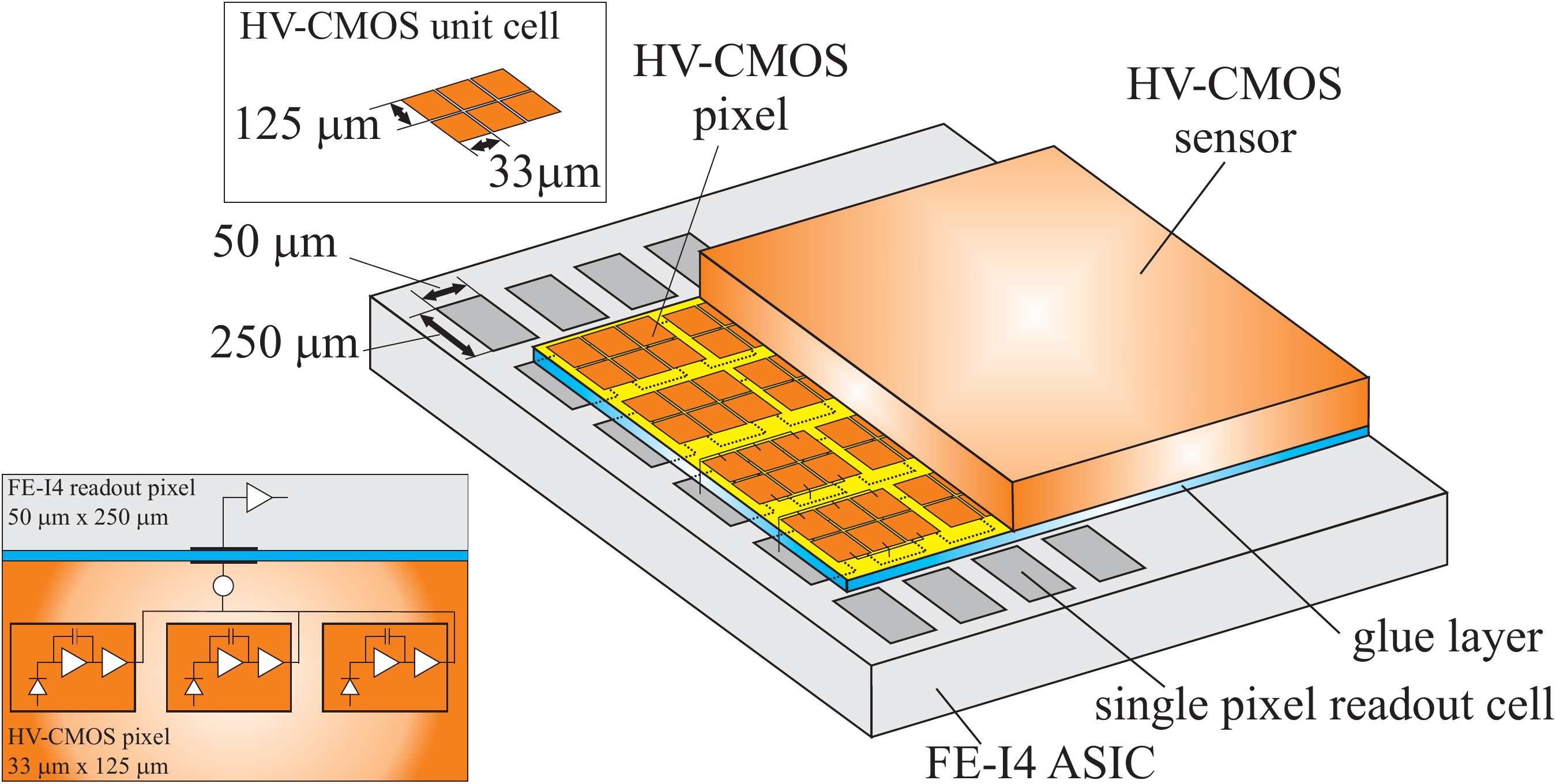}
\caption{Schematic representation (not in scale) of the HV-CMOS to FE-I4B connections. The bottom-left inset shows how three HV-CMOS pixels (forming a so-called {\em macro-pixel}) are capacitively coupled to a single FE-I4B readout pixel.}
\label{fig:ccpd-fei4-3D}
\end{figure}

As the FE-I4B pixel size is \mbox{$250 \times 50$~\micron\squared}, the unit cell, made by six HV-CMOS pixels, corresponds to two FE-I4 pixels (see Fig.~\ref{fig:ccpd-fei4-3D}). In this respect, we define \emph{macro-pixel} as the set of three HV-CMOS single pixels (half the unit-cell) connected to a single FE-I4B readout pixel. For simplicity of operation in the testbeam, the signal amplitude of the discriminator output of each of these three sub-pixels were set to the same voltage level. 

\subsection{Charge calibration}
\label{sec:charge_calibration}

In order to perform a voltage-to-charge calibration, a setup with a $\rm{^{55}Fe}$ radioactive source was used. The most intense emission of this X-ray source is at 5.9 keV ($K_\alpha$ peak). Photons are absorbed in the silicon bulk creating photoelectrons that can then be collected. The $\rm{^{55}Fe}$ spectrum was recorded by monitoring with an oscilloscope the output signal of a single pixel after the amplification stage. The measurement was performed at room temperature with the sensor  biased at 80 V. In order to estimate the intrinsic noise level of the setup, a first acquisition without the source was performed. The threshold of the oscilloscope was then set at five times the r.m.s. of the (Gaussian-distributed) noise distribution. By determining the position of the $K_\alpha$ peak and assuming 3.65 eV as the ionization potential for silicon at room temperature, the resulting calibration factor is $\sim$8.6 $\rm{e^-}$/mV. The nominal operating threshold of the sensor corresponds to $\sim$600 $\rm{e^-}$.

\subsection{Threshold tuning}
\label{sec:dut_threshold}

\begin{figure}[!tbp]
\centering
\subfloat[]{\label{fig:car}
\includegraphics[width=0.30\textwidth]{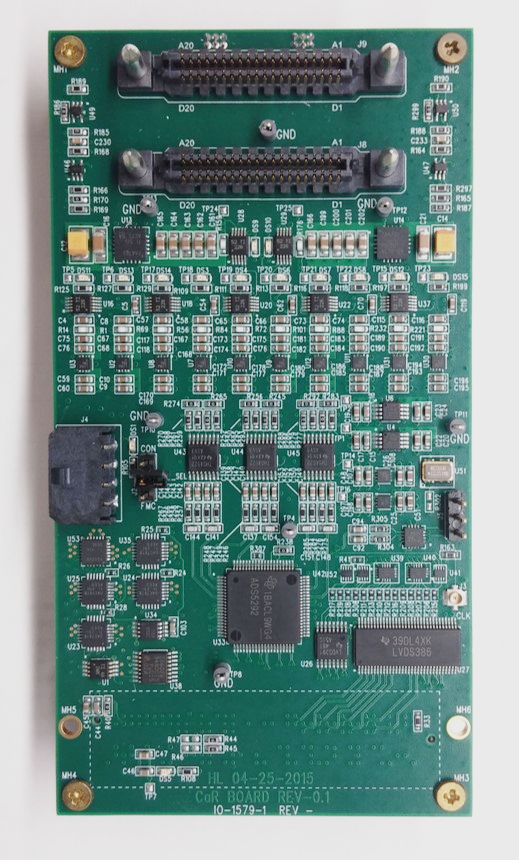}}
\hspace{2cm}
\subfloat[]{\label{fig:car_fei4}
\includegraphics[width=0.53\textwidth]{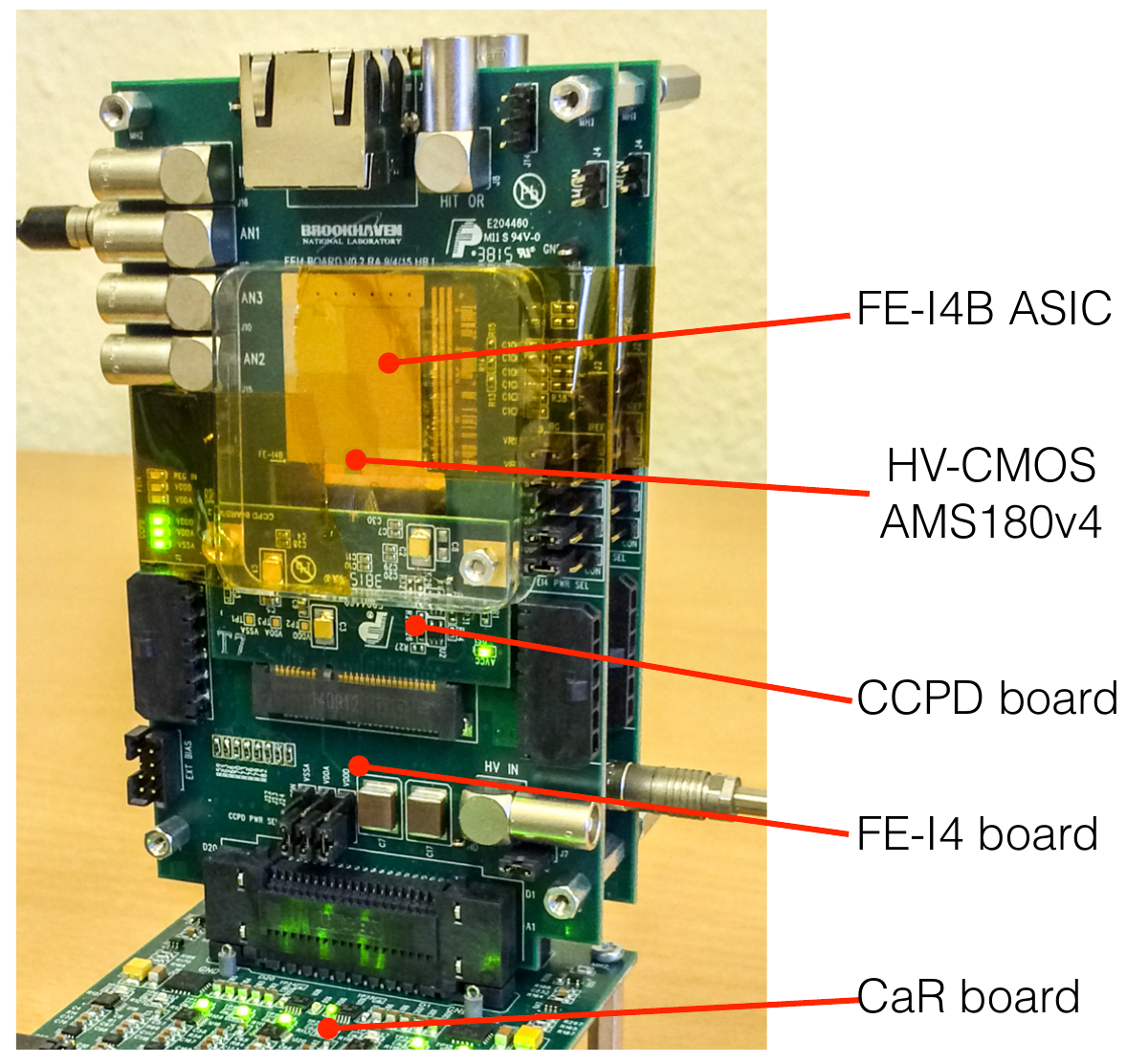}}
\caption{(a) Top-side of the CaR board. (b) The FE-I4 board with a CCPD board attached. In the configuration shown, two FE-I4 boards are connected to a single CaR board.}
\label{fig:caribou}   
\end{figure}

The powering, configuration and control of the AMS180v4 prototype was performed using the CaRIBOu~\cite{caribou} ({\em Control Readout Itk BOard}) test system. CaRIBOu is a modular readout system developed by the Brookhaven National Laboratory and the University of Geneva groups to characterize HV-CMOS sensor prototypes in  laboratory and testbeam measurements. A Xilinx Z0706 evaluation board~\cite{xilinx} with embedded ARM processor, based on the Zynq\textregistered-7000 All Programmable SoC, acts as main DAQ board. The system comprises various additional custom-designed and inter-connected boards (see Fig.~\ref{fig:caribou}):
\begin{itemize}
\item The CaR ({\em Control and Readout}) board provides powering rails, adjustable bias DACs, CMOS level translators, LVDS and clock signals, a digital injection circuitry and ADC channels. It is connected to the Xilinx Z0706 board via a FPGA Mezzanine Card (FMC) to a Very-High-Density Cable Interconnect (VHDCI) interface.
\item The FE-I4 board contains the external high-voltage input for the sensor bias, the bonding pads for the FE-I4B ASIC and the required circuits for both the HV-CMOS and the FE-I4B ASIC. 
\item The CCPD board simply contains passive SMD components and the bonding pads for the HV-CMOS sensor. The CCPD board is attached to the FE-I4 board via a PCI Express mini-connector.
\end{itemize}
While the Z0706 and CaR boards were used to configure and control the HV-CMOS sensor, the glued FE-I4 chip was readout by the same RCE system as the telescope. A binary tuning procedure was implemented to minimize the threshold spread among all channels of the sensor. The tuning method consists in selecting a target threshold and then finding the optimal TDAC setting for each pixel so that the resulting threshold is close to the target within a defined tolerance. At each threshold scan, the TDAC is varied by a single unit step. A Gaussian fit to the threshold distribution obtained in the DUT after tuning resulted in a mean and sigma of  \mbox{607 $\rm{e^-}$} and \mbox{73 $\rm{e^-}$}, respectively.

\section{Off-line analysis}
\label{sec:offline}

The raw data from all six telescope planes and from the DUT are stored in ROOT~\cite{root} format for off-line processing. The software analysis framework is based on the Judith~\cite{judith} program, a C++ software package originally developed for the Kartel telescope (composed of six MIMOSA-26 pixel sensors~\cite{mimosa}). The Judith package has been adapted for the FE-I4 telescope configuration and module geometries.

The off-line analysis compares the response of the DUT to tracks reconstructed with the beam telescope. The internal alignment of the different telescope planes is performed using an iterative algorithm that minimizes the track residuals, {\em i.e.} the difference between the hit position measured in the plane being aligned and the predicted hit position from the track extrapolation into that plane. Noisy pixels are masked at software level before clustering. The first telescope plane is kept fixed at its nominal position as an anchor reference to define the $X$- and $Y$-axis. The track reconstruction algorithm recursively searches for clusters in consecutive planes within a limited solid angle to deal with multiple scattering. A straight line fit is then performed to the different cluster positions. The alignment algorithm corrects for the translations in the transverse plane and for the in-plane rotation around the $Z$-axis. Results from the telescope alignment can be found in~\cite{telescopePaper}. Upper limits on the spatial resolution of the telescope were found to be 11.7~\micron\ and 8.3~\micron\ at the DUT position.

\begin{figure}[!tbp]
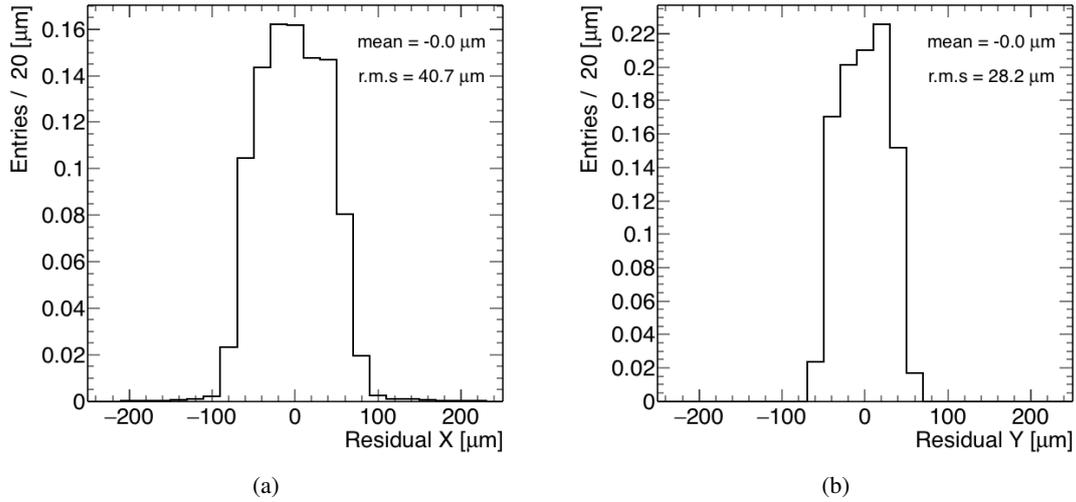

\centering
\subfloat[]{\label{fig:dutResX}
\includegraphics[width=0.45\textwidth]{figure5a.pdf}}
\hspace{0.5cm}
\subfloat[]{\label{fig:dutResY}
\includegraphics[width=0.45\textwidth]{figure5b.pdf}}
\caption{Normalized residual distributions of the DUT along the $X$ (a) and $Y$ (b) directions.}
\label{fig:dutRes}   
\end{figure}

After the internal alignment of the FE-I4 telescope, reconstructed tracks are used to align the DUT. Due to the small size of the HV-CMOS sensor there was not enough sensitivity to correct for the in-plane rotation, so in this case only the two translations were considered. Single track events were selected. Tracks were required to have one hit in each of the six telescope planes, a good $\chi^2$ from the track fit and small track gradients with respect to the beam axis. Figure~\ref{fig:dutRes} shows the unbiased residual distributions at the DUT after alignment along the two measuring directions. The \mbox{r.m.s. of the residual distributions} are in reasonable agreement with the expected intrinsic resolutions of the \emph{macro-pixel} along the $X$ ($\sigma_{x,\text{exp}}$) and $Y$-directions ($\sigma_{y,\text{exp}}$), \mbox{$\sigma_{x,\text{exp}}=125~\micron\,/\sqrt{12}=36.1~\micron$} and \mbox{$\sigma_{y,\text{exp}}=100~\micron\,/\sqrt{12}=28.9~\micron$}, respectively.  

	\section{Results}
\label{sec:results}

\subsection{Cluster size distributions}

\begin{figure}[!tbp]
\centering
\includegraphics[width=0.45\textwidth]{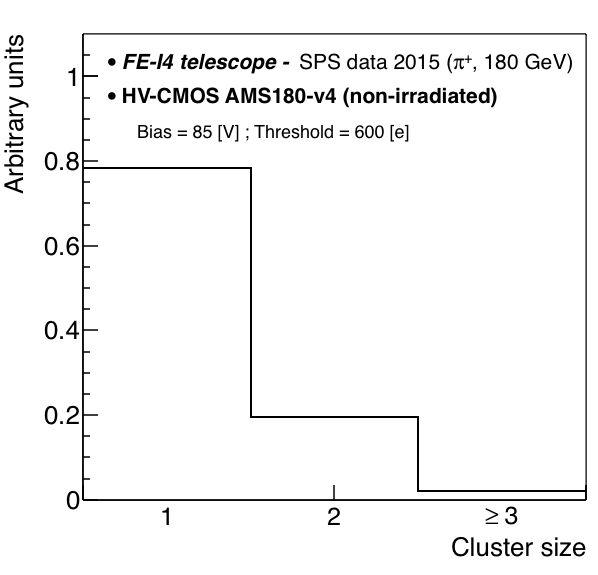}
\caption{Cluster size distribution for normal incidence tracks and nominal operating conditions. The distribution has been normalized to unity.}
\label{fig:cs}
\end{figure}

The cluster size is the number of neighbouring pixels (along both measuring directions) with a signal above threshold. It is an effective measurement of charge sharing. Figure~\ref{fig:cs} shows the cluster size distribution for the DUT at nominal operating conditions (85 V bias-voltage and 600 $\rm{e^-}$ threshold). At the SPS testbeam the track incidence was almost perpendicular to the detector planes. In $\sim$78\% of the events single-pixel clusters were recorded, while clusters formed by two pixels were observed in $\sim$20\% of the cases. The occurrence of clusters with sizes of three or more pixels was less than 2\% and thus can be safely neglected in what follows. Among the possible mechanisms leading to signal leakage into neighboring pixels ({\em e.g.} cross-talk between channels from capacitive coupling, diffusion of charge carriers, $\delta$-electrons), diffusion is the dominant one. This diffusion component is only relevant for unirradiated sensors, as it is largely suppressed by charge trapping  for fluences typically above $10^{15}\,\rm{n_{eq}/cm^2}$, as measured on irradiated HV-CMOS prototypes with an edge-TCT technique~\cite{marcos} and with testbeam data~\cite{javier}.

\begin{figure}[!tbp]
\centering
\subfloat[]{\label{fig:timing_allCS}
\includegraphics[width=0.45\textwidth]{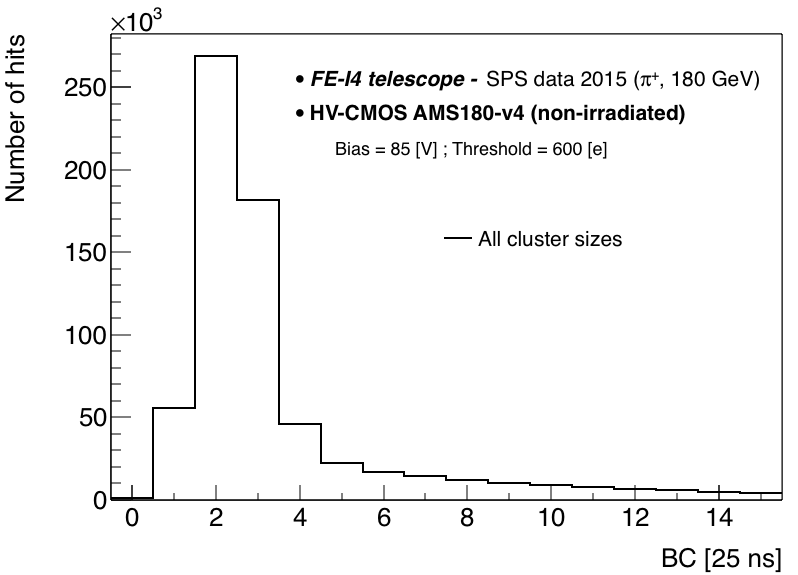}}\\
\subfloat[]{\label{fig:timingCS_1}
\includegraphics[width=0.45\textwidth]{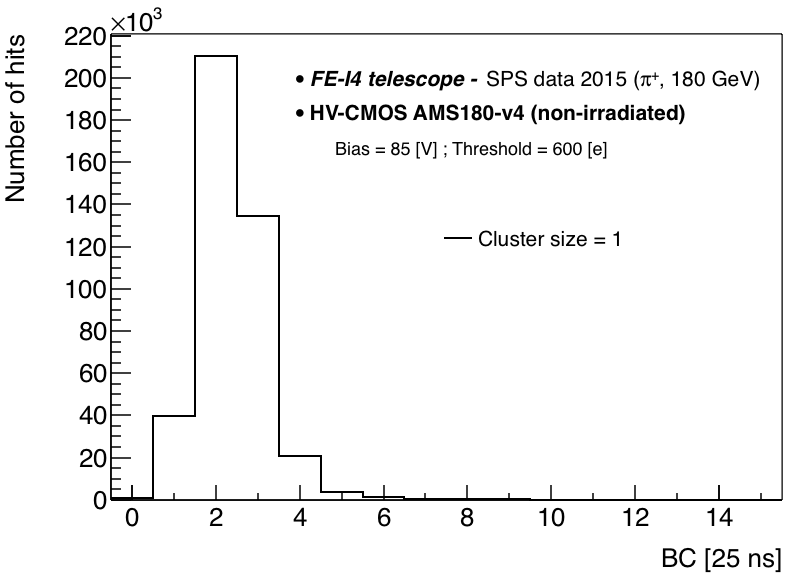}}
\hspace{0.5cm}
\subfloat[]{\label{fig:timingCS_2}
\includegraphics[width=0.45\textwidth]{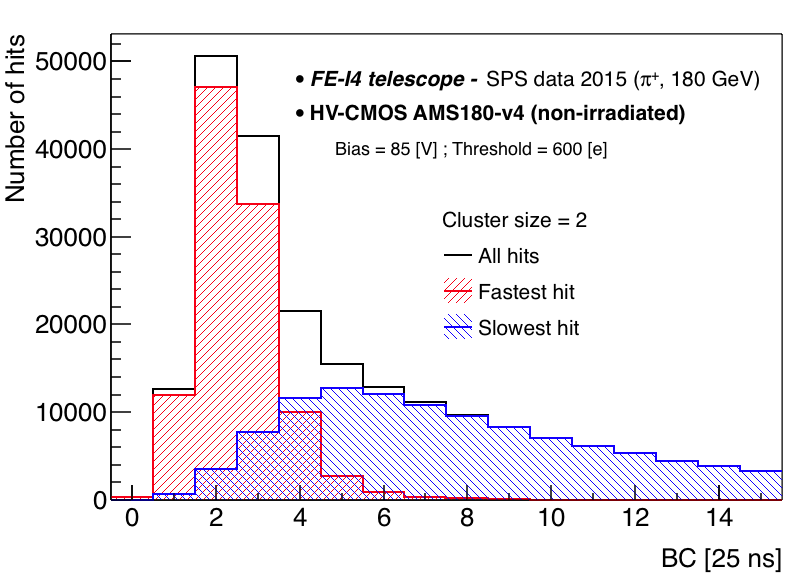}}
\caption{Timing distribution of the DUT hits in 25 ns bins for all cluster sizes (a), cluster size one (b) and cluster size two (c). In (a), all pixels are shown. In (c), the distribution of the measured time of each of the two pixels in the cluster is shown.}
\label{fig:timingCS}   
\end{figure}

Figure~\ref{fig:timing_allCS} shows the timing distribution ({\em i.e.} the timestamp difference of the recorded hit with respect to the trigger in units of BC, see Sec.~\ref{sec:telescope}) for all cluster sizes. The distribution shows a long tail of late (slow) hits. This tail is mainly due to a known timewalk effect in this HV-CMOS sensor design~\cite{javier}, so that low amplitude signals such as resulting from diffusion have a larger timing because of the slower discriminator rise time. In Figs.~\ref{fig:timingCS_1} and ~\ref{fig:timingCS_2} we investigated the contribution from clusters of size one and two, respectively. In the latter case, the timing of both hits in the cluster are shown. The timing distribution of clusters of size one is very similar to that of the fastest hit of the two-pixel clusters. The slowest signals in the tail of the distribution of Fig.~\ref{fig:timing_allCS} thus correspond to the diffused charge in large multiplicity clusters (charge sharing). In order to focus on the properties of the charge collected by drift, and to limit the effects of the known timewalk present in this version of the HV-CMOS sensor, only results for the fastest hit are shown in the following.

\subsection{Efficiency}

\begin{figure}[tbp]
\centering
\includegraphics[width=0.6\textwidth]{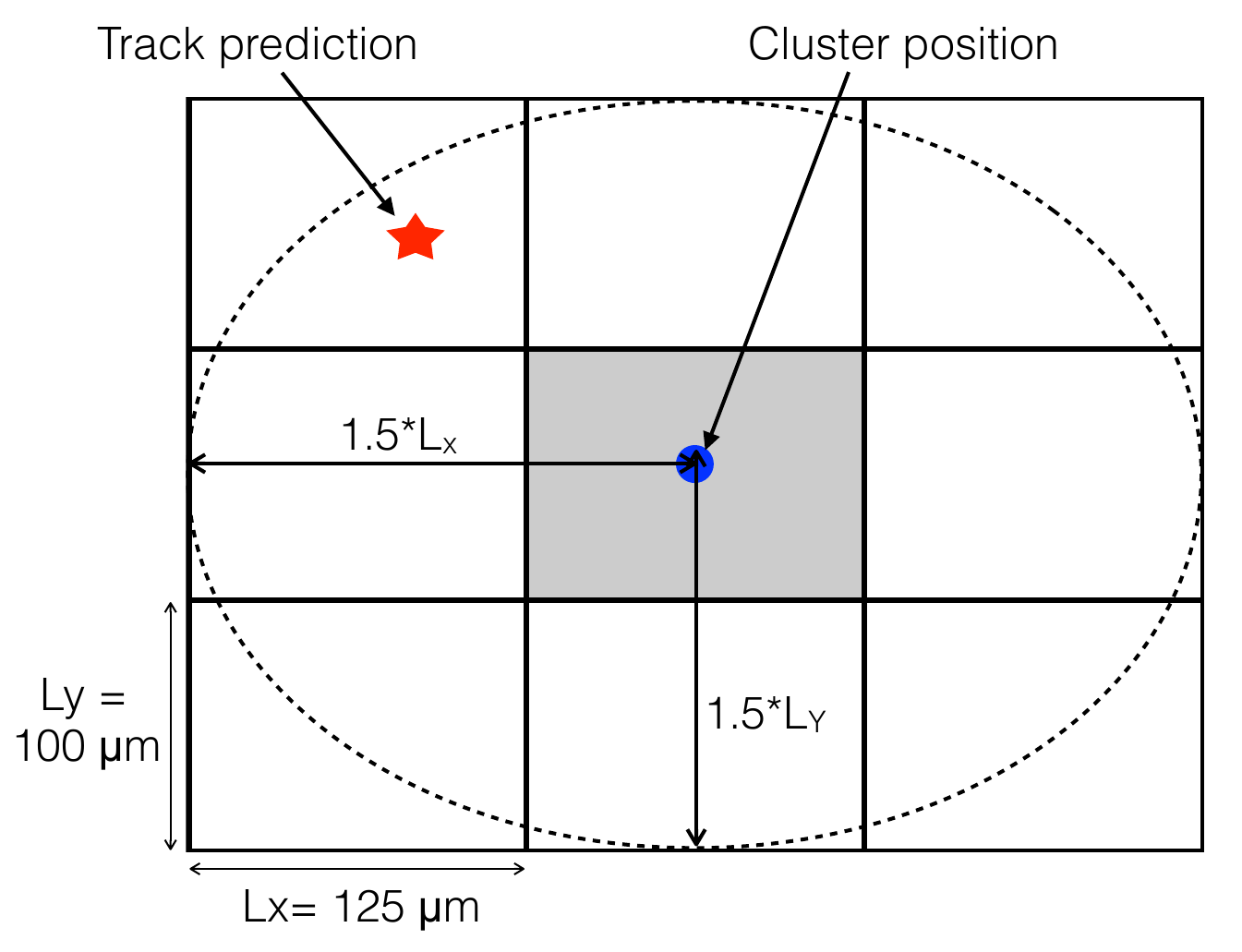}
\caption{Definition of the area used to calculate the efficiency. Each \mbox{125~\micron\ $\times$ 100~\micron} \emph{macro-pixel} (gray area) is formed by three single HV-CMOS \mbox{125~\micron\ $\times$ 33~\micron} pixels.}
\label{fig:ellipse}
\end{figure}

\begin{figure}[!tbp]
\centering
\includegraphics[width=0.75\textwidth]{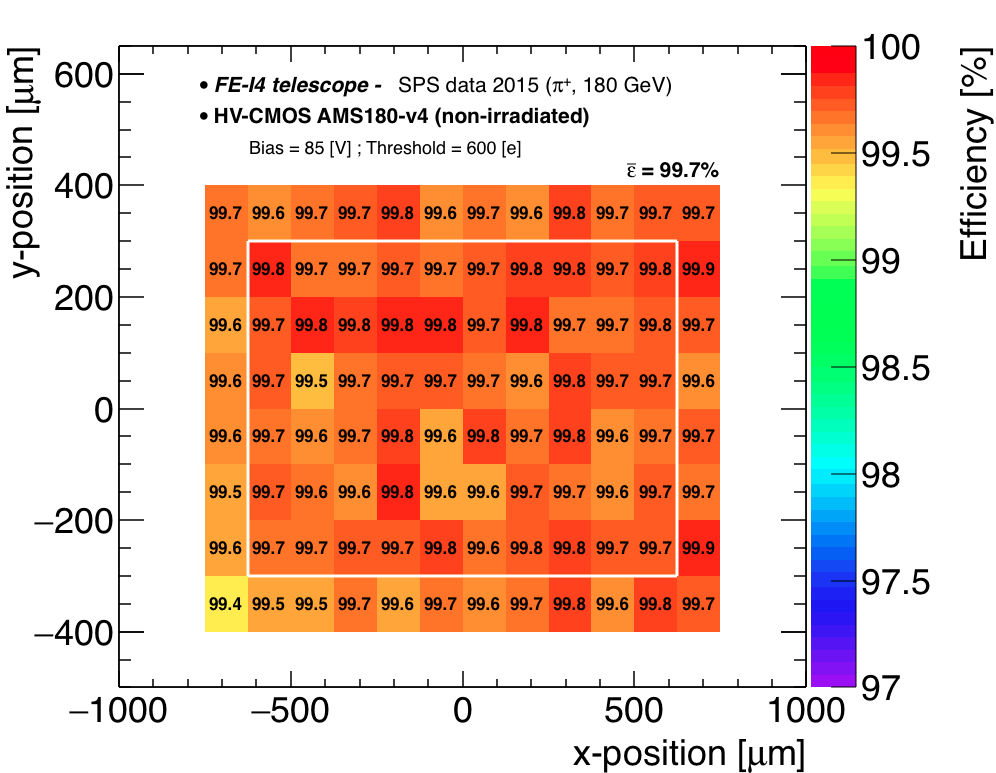}
\caption{Efficiency map of the AMS180v4 sensor. For the computation of the average global efficiency, $\bar\varepsilon=99.7\%$, the pixels inside the white box are used.}
\label{fig:globeff}
\end{figure}

One of the most important parameters to be studied of the AMS180v4 prototype is the tracking efficiency. For this measurement, a detector plane is considered to be efficient if the interpolated track position is located within an ellipse centered around the measured cluster with semi-axes being one and a half times the size of the \emph{macro-pixel} (see Fig.~\ref{fig:ellipse}). The lengths of the ellipse semi-axes correpond to five times the standard deviation of the residual distributions (see Fig.~\ref{fig:dutRes}). The cluster position is computed as the geometrical center of the \emph{macro-pixel}. Single track events and the track quality cuts as defined in Sec.~\ref{sec:offline} are used. The interpolated track position is required to be inside the DUT sensitive area, otherwise the event is discarded. The efficiency of a \emph{macro-pixel} is thus simply computed as \mbox{$\varepsilon = N_\text{eff}/N_\text{tot}$}, where $N_\text{eff}$ is the total number of efficient events in the pixel as just defined, and $N_\text{tot}$ is the total number of selected events. Inefficient events are taken into account (these are events in which for a valid track interpolation at the DUT position, no hit has been measured in the DUT). Fig.~\ref{fig:globeff} shows the resulting efficiency map for a bias voltage of 85 V and a threshold of 600 $\rm{e^-}$. Excluding the outermost columns and rows to avoid edge-effects due to the finite telescope resolution, an average global efficiency of $\bar\varepsilon=99.7\%$ is obtained.

\begin{figure}[!tbp]
\centering
\subfloat[]{\label{fig:inPixel_all}
\includegraphics[width=0.45\textwidth]{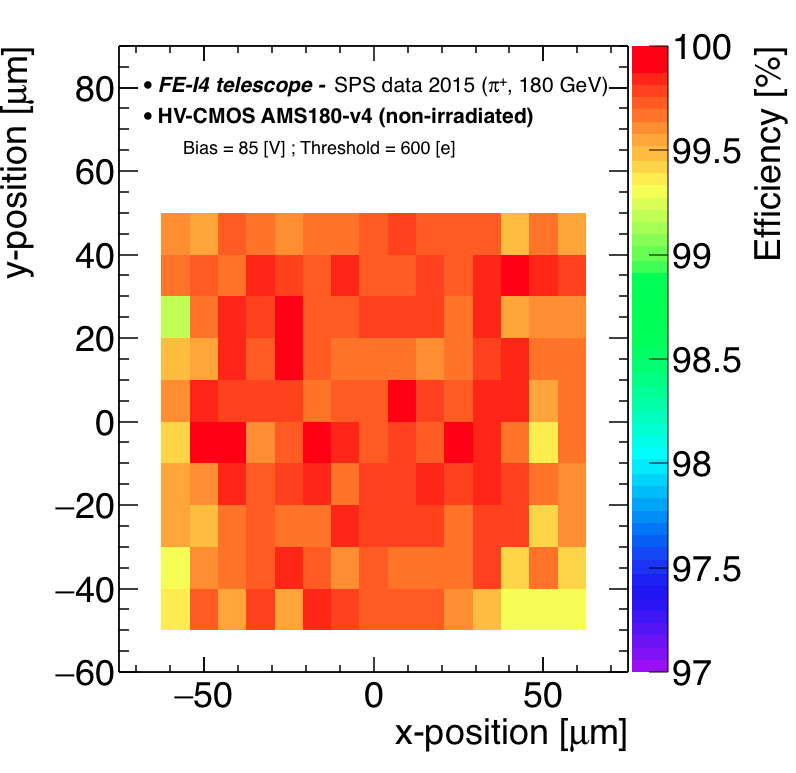}}
\hspace{1cm}
\subfloat[]{\label{fig:inPixel_1}
\includegraphics[width=0.45\textwidth]{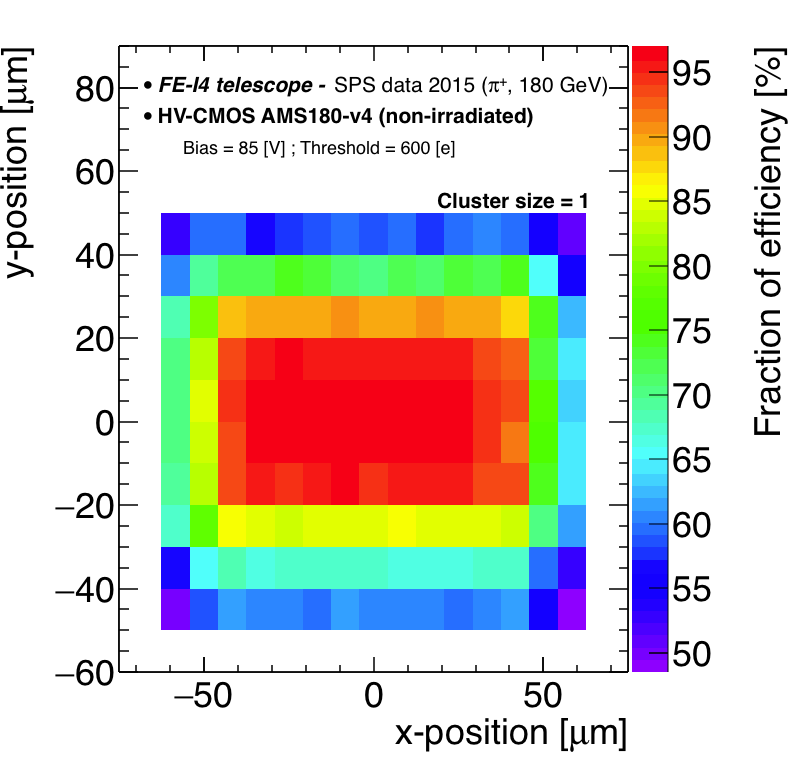}}\\
\subfloat[]{\label{fig:inPixel_2}
\includegraphics[width=0.45\textwidth]{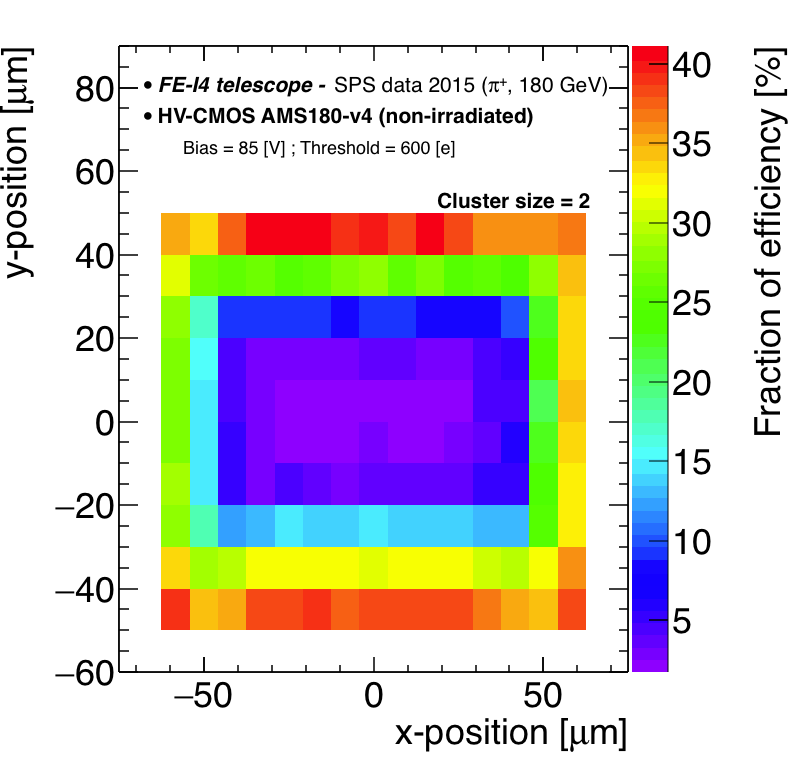}}
\hspace{1cm}
\subfloat[]{\label{fig:inPixel_geq3}
\includegraphics[width=0.45\textwidth]{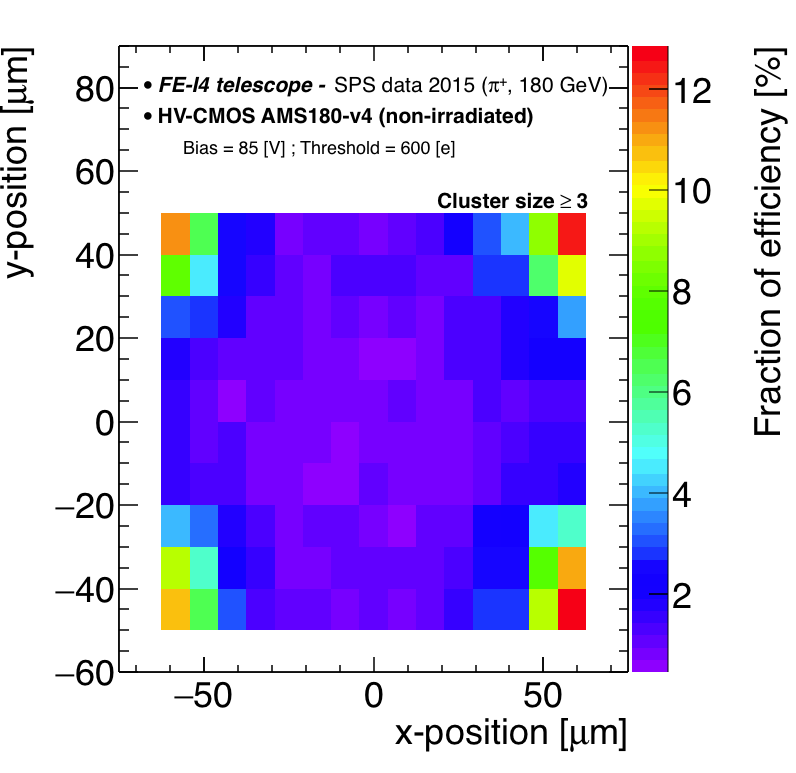}}
\caption{(a) Efficiency inside a \emph{macro-pixel} of $125~\micron\times 100~\micron$, corresponding to three single HV-CMOS pixels connected to a FE-I4B readout cell. (b-d) Relative contributions to the distribution shown in (a) from pixel clusters of different sizes.}
\label{fig:inPixel}   
\end{figure}

Thanks to the pointing resolution of the FE-I4 telescope~\cite{telescopePaper} a sub-pixel resolution was achieved. Fig.~\ref{fig:inPixel_all} shows the  efficiency map after merging all {\em STime} pixels into a single histogram representing a \emph{macro-pixel} ($125\times 100$~\micron\squared, corresponding to three HV-CMOS pixels connected to a single FE-I4B readout channel). Figs.~\ref{fig:inPixel_1} to~\ref{fig:inPixel_geq3} show the relative contributions to the efficiency of different cluster sizes. The high efficiency in the central region of Fig.~\ref{fig:inPixel_all} results dominantly from events where the cluster size is exactly one ($\sim$78\% of the cases, see Fig.~\ref{fig:cs}). Events with a cluster size of two ($\sim$ 20\% of the cases) contribute mainly to the edge regions, as expected from charge sharing between neighbouring pixels. The less common events in which the cluster contains three or more pixels are localized at the four corner regions. 

\begin{figure}[tbp]
\centering
\includegraphics[width=0.6\textwidth]{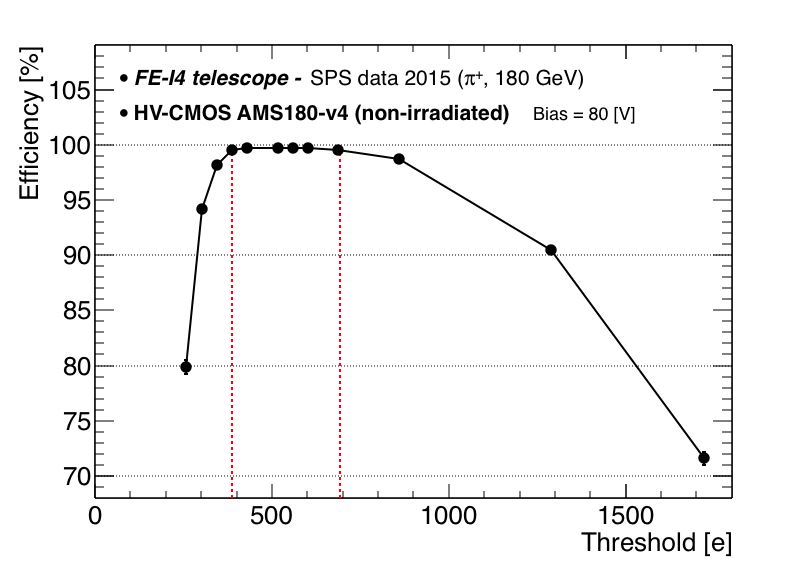}
\vspace{-0.2cm}
\caption{Efficiency as a function of threshold. The two vertical dashed-lines indicate the operating region for which the DUT efficiency is larger than 99.5\%.}
\label{fig:thScan}
\end{figure}

The efficiency as a function of the discriminator threshold is shown in Fig.~\ref{fig:thScan} for a bias voltage of 80 V. As expected, the efficiency drops in the low-threshold (noise dominated) and high-threshold (low signal) regions. The region for which the efficiency is higher than 99.5\% is between 385 and 690 $\rm{e^-}$.

\begin{figure}[!htbp]
\centering
\subfloat[]{\label{fig:biasScan}
\includegraphics[width=0.57\textwidth]{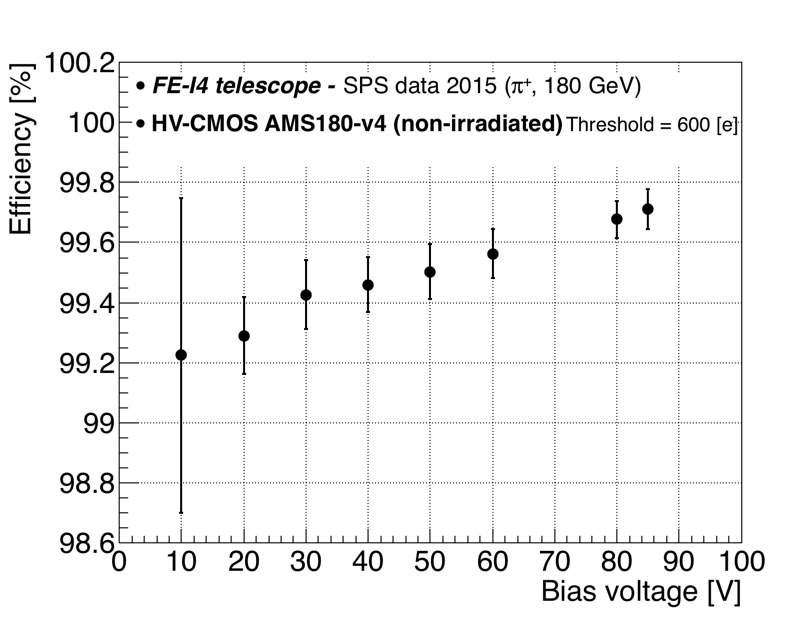}}
\subfloat[]{\label{fig:inPixel_20V}
\includegraphics[width=0.44\textwidth]{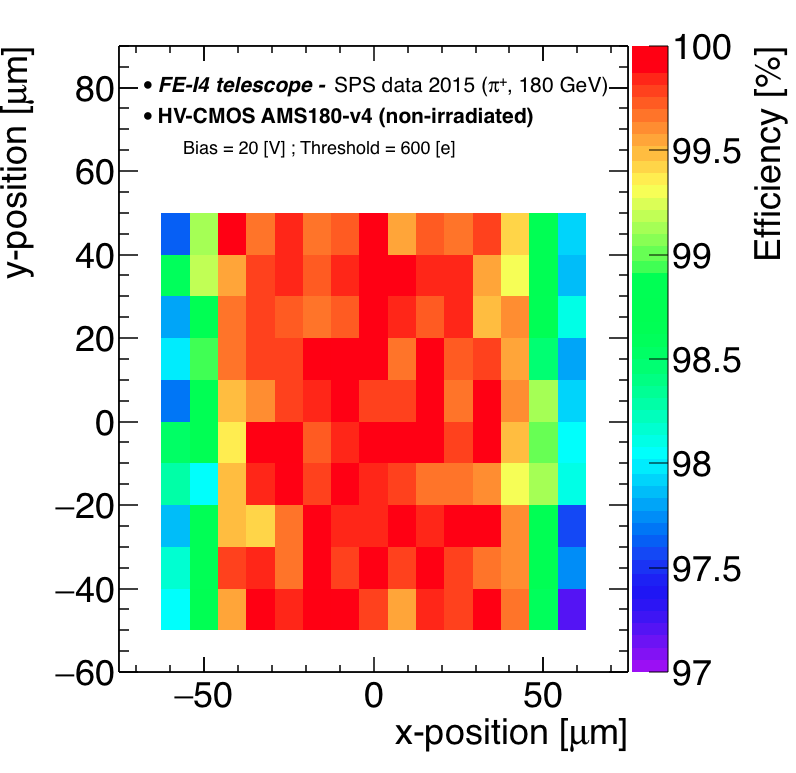}}
\vspace{-0.1cm}
\caption{(a) Efficiency as a function of bias voltage. (b) In-pixel efficiency for 20 V bias voltage. The reduced efficiency on the vertical edges is due to a higher charge sharing as from the finer pixel granularity along the vertical direction (the \emph{macro-pixel} shown corresponds to three $125~\micron\times 33~\micron$ pixels).}
\end{figure}

The efficiency as a function of the sensor bias voltage is shown in Fig.~\ref{fig:biasScan} for a threshold of 600 $\rm{e^-}$. The large statistical uncertainty of the first data point (10 V bias voltage) is due to the fact that, accidentally for this particular run, the ROI region was not set. Fig.~\ref{fig:inPixel_20V} shows the in-pixel efficiency for a bias of \mbox{20 V} (to be compared with Fig.~\ref{fig:inPixel_all} where the bias voltage was \mbox{85 V} at the same threshold). The lower depletion region is source of inefficiencies at the pixel edges. Overall, the relatively high efficiency at lowest bias voltages indicates that, in addition to drift, part of the charge is collected by diffusion. 

\begin{figure}[!tbp]
\centering
\subfloat[]{\label{fig:timing}
\includegraphics[width=0.48\textwidth]{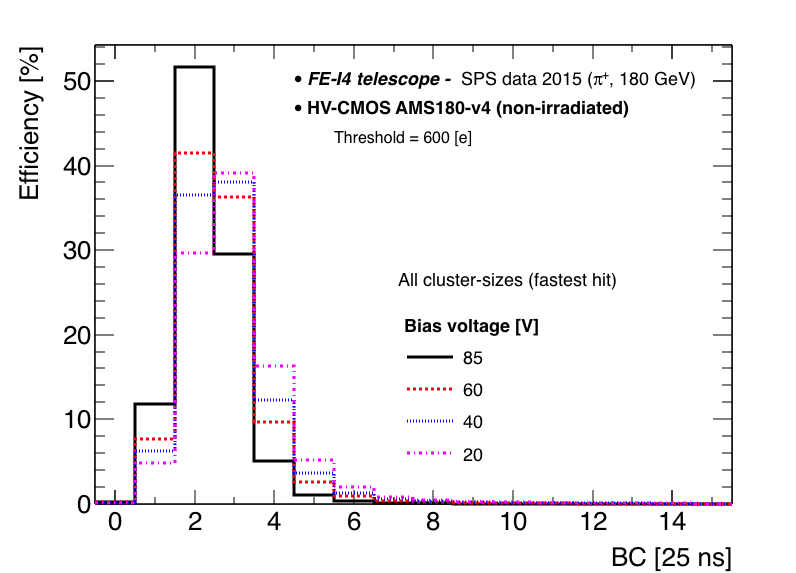}}
\subfloat[]{\label{fig:timingCumul}
\includegraphics[width=0.48\textwidth]{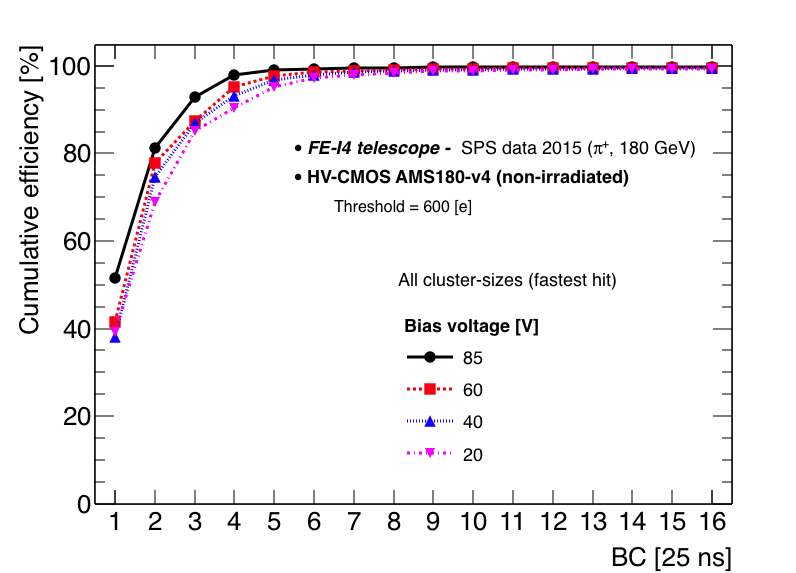}}
\caption{(a) Efficiency and (b) cumulative efficiency in bins of 25 ns for different sensor bias voltages. For cluster sizes of two or more, only the fastest hit is shown. The first data point shown in (b) corresponds to the maxima of the distributions shown in (a).}
\end{figure}

The efficiency has been evaluated in each of the sixteen 25 ns time intervals used for the readout of the DUT. Figure~\ref{fig:timing} shows the efficiency per time-bin for different bias voltages for the fastest hit in the cluster. The shape of the distributions are similar to that already shown in Fig.~\ref{fig:timingCS_1}. Due to some existing delays (\emph{e.g.} different cable lenghts, clock-jitter) between the trigger and the  DUT readout, it is assumed that the time-bin for which the maximum is reached corresponds to that of the passage of the incident particle. Most particles are detected in a two-bin interval. For increasing bias voltages the depletion region enlarges and the maximum efficiency increases. 
Fig.~\ref{fig:timingCumul} shows the cumulative efficiency as a function of 25 ns time-bins. The distribution has been computed with respect to the consecutive maxima in Fig.~\ref{fig:timing}. At 85 V bias voltage, an efficiency higher than 90\% is achieved after three BCs.

\begin{figure}[htbp]
\centering
\includegraphics[width=0.55\textwidth]{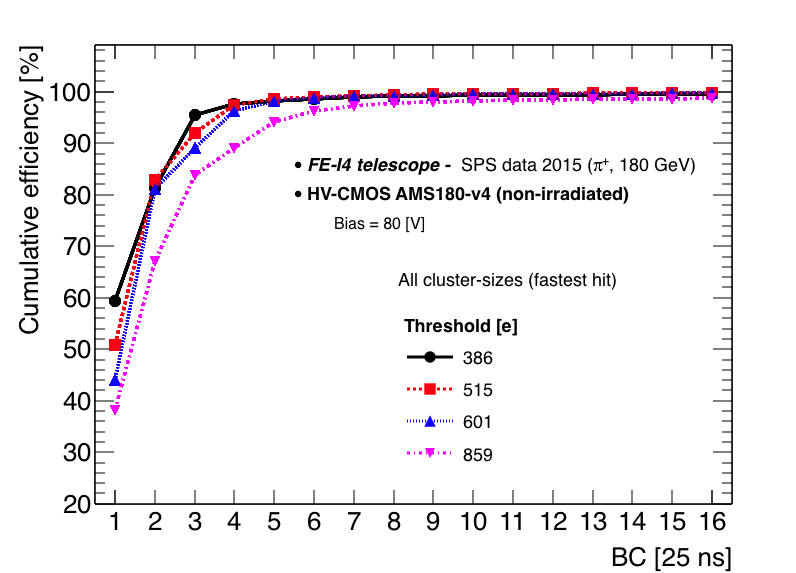}
\caption{Cumulative efficiency in bins of 25 ns for different thresholds. For cluster sizes of two or more, only the fastest hit is shown. }
\label{fig:cumul_thScan}
\end{figure}

Figure~\ref{fig:cumul_thScan} demonstrates the effect of the threshold on the timing distribution of the DUT. As the threshold is lowered, it can be observed that the dispersion of the time of arrival of the hits is reduced. This is further evidence of the large variation of the rise time of the signal. In the case where the rise time of the signal would be the same for small and large signal, a reduction of threshold would only yield an increase in efficiency. However it is observed here that the improvement of the timing distribution spread is occurring even when the threshold is reduced within the plateau of efficiency featured in Fig.~\ref{fig:thScan}. 

\section{Conclusions}
\label{sec:conclusions}

A HV-CMOS sensor prototype produced in the AMS \mbox{180 nm} technology has been tested in November 2015 at the H8 beamline of the CERN SPS. A FE-I4B ASIC has been glued to the HV-CMOS sensor to read the signals via capacitive coupling. For the first time, the CaRIBOu system has been used to power, configure and control the AMS180v4 chip. A threshold uniformization using a binary tuning algorithm yielded a final threshold dispersion of $\sim$70~$\rm{e^-}$ about a mean of $\sim$610~$\rm{e^-}$. The FE-I4 beam telescope has been used to reconstruct the tracks to evaluate the performance of the HV-CMOS prototype. An average global efficiency of 99.7\% has been measured for a bias voltage of 85 V and an operating threshold of 600 $\rm{e^-}$. It has been measured that the sensor can be operated with an efficiency higher than 99.5\% within a threshold range between $\sim$400 and $\sim$700 $\rm{e^-}$. The timing performance of this prototype is not yet sufficient to be operated in a HL-LHC experiment as an efficiency higher than 90\% is measured within a three 25 ns BC window. New versions of the AMS chip have already been produced in both 350 and 180 nm technologies to improve the time response of the sensor. 
\acknowledgments

We gratefully acknowledge the support by the CERN PS and SPS instrumentation team. 
We thank Allan Clark for a careful reading of the manuscript. The research presented in this paper was supported by the SNSF grants 200020\_156083, 20FL20\_160474 and 200020\_163402.

\end{document}